\documentclass[12pt]{article}
\usepackage{epsfig}

\begin{document}

\title{Role of unphysical solution in nucleon QCD sum rules
}
\author{E. G. Drukarev, M. G. Ryskin, V. A. Sadovnikova,\\
Petersburg Nuclear Physics Institute\\
Gatchina, St. Petersburg 188300, Russia
}
\date{}
\maketitle

\begin{abstract}
We show that at certain values of QCD condensates the nucleon QCD sum
rules with ``pole+continuum" model for the hadron spectrum obtain an
unphysical solution. This provides constrains for the values of
condensates to be consistent with existence of a physical
solutions. The constrains become much weaker if the radiative
corrections are included perturbatively. We demonstrate that the most
important dependence of nucleon mass on the quark scalar condensate
becomes less pronounced under factorization assumption for the four-quark
and six-quark condensates.

\end{abstract}

\section{Introduction}

The QCD sum rules invented by Shifman {\em et~al.} \cite{1} enable to
express vacuum characteristics of hadrons in terms of the vacuum
expectation values of QCD operators. This approach was employed
to description of nucleons in \cite{2} and \cite{2a,2b}. The improved
analysis was  presented later
in \cite{2c,2d}. The method was used also for description of delta-isobars
\cite{2c,2d} and of the baryons containing heavier quarks
\cite{3a,3b}. Further
applications of the vacuum nucleon QCD sum rules are reviewed in \cite{3}.
The method was expanded also for the cases of finite temperatures~\cite{4}
and densities~\cite{5}.

The main tool of the vacuum QCD sum rules for a hadron is the
dispersion relation for polarization operator
\begin{equation}
\Pi(q^2)\ =\ i\int d^4xe^{i(qx)}\,\langle0|Tj(x)\bar j(0)|0\rangle\,,
\end{equation}
with $j(x)$ a local operator ("current") carrying the quantum
numbers of the hadron. It is nucleon (proton) in our case. The dispersion
relation is considered at large values of $|q^2|(q^2<0)$ where
$\Pi(q^2)$ can be represented as a power series in $q^{-2}$, with
the vacuum expectation values of QCD operators as the coefficients
of the expansion. This is known as operator power expansion (OPE)
\cite{6}.

Recall the main milestones of the QCD sum rules analysis. Following
\cite{2} one can write dispersion relations
\begin{equation}
\Pi^i(q^2)\ =\ \frac1\pi \int \frac{\mbox{Im
}\Pi^i(k^2)}{k^2-q^2}\,dk^2
\end{equation}
$(i=q,I)$ for the ingredients $\Pi^i(q)$ of the polarization operator
\begin{equation}
\Pi(q)\ =\ q_\mu\gamma^\mu\Pi^q+I\Pi^I(q^2)\,.
\end{equation}
We equal the OPE of the left-hand side (LHS) of Eq.~(2) to the
contribution of the observable hadrons to its right-hand side (RHS).
The latter is usually approximated by the ``pole + continuum" model in
which the lowest pole is written exactly while the other states are
approximated by continuum:
\begin{equation}
\frac1\pi\mbox{ Im }\Pi^i(k^2)\ =\ \lambda^2_N\delta(k^2-m^2)+
\theta(k^2-W^2)f^i(k^2)\,.
\end{equation}
Of course, the "pole + continuum" model is reasonable
only if the contribution of the pole exceeds that of the continuum.

Behavior of both sides of Eq.~(2) prompts the choice \cite{1}
$$
f^i(k^2)\ =\ \frac1\pi\mbox{ Im }\Pi^{i(\rm OPE)}(k^2)\,.
$$
Note that in this approach the continuum threshold $W^2$ does not
coincide with the physical continuum threshold. Thus the position of
the lowest pole $m$, its residue $\lambda^2_N$ and the model continuum
threshold $W^2$ are the unknowns which are expected to be determined by
the QCD sum rules equations. The standard next step is the Borel
transform, after which Eqs.(2) take the form
\begin{equation}
{\cal L}^q(M^2)=R^q(M^2); \quad {\cal L}^I(M^2)=R^I(M^2).
\end{equation}
Here ${\cal L}^i(R^i)$ are the Borel transforms of the LHS (RHS) of
Eq.~(2), $M^2$ is the Borel mass.
This approach provided good results for the nucleon mass and
for the other nucleon parameters \cite{3}.

Note that both ${\cal L}^i(M^2)$ and $R^i(M^2)$ are calculated in framework of certain models.
The OPE expansion for ${\cal L}^i(M^2)$ is increasingly true at large values of $M^2$.
The ``pole +
continuum" for model for $R^i(M^2)$is increasingly true at small values of $M^2$.
Important assumption is that there is a region of intermediate
values of $M^2$ where both approximations work and reproduce to some
extend the true (unknown) function of $M^2$. Thus our task is to find the interval of
the values of $M^2$, where the functions ${\cal L}^i(M^2)$ can be approximated by the
functions $R^i(M^2)$ and to find the set of parameters $m, \lambda_N^2$ and $W^2$ which insure
the most accurate approximation of ${\cal L}^i(M^2)$ by $R^i(M^2)$.
The set of values of parameters $m,\lambda^2_N$ and $W^2$, which
minimize the function
\begin{equation}
\chi^2(m, \lambda_N^2, W^2)\ =\ \sum_j\sum_{i=q,I} \left(\frac{{\cal
L}^i(M^2_j)-R^i(M^2_j)}{{\cal L}^i(M^2_j)}\right)^2
\end{equation}
will be referred to as a solution of the sum rules equations.

Note that it is important to obtain ``duality" between the LHS and
RHS of Eq.(5) in some interval of the values of $M^2$, but not at
certain point $M^2_j$. Therefore we will look for the three
unknown parameters simultaneously.

Both the interval of the values of the Borel mass ("Borel window")
and the solution of the sum rules depend on the form of the
proton nucleon current $j(x)$, which is not determined in an unique way.
The general form is \cite{2,6a}
\begin{equation}
j(x;t)=j_1(x)+tj_2(x),
\end{equation}
with
$$ j_1(x)= \varepsilon_{abc}[u_a^T(x)Cd_b(x)]\gamma_5u_c(x); \quad
j_2(x)= \varepsilon_{abc}[u_a^T(x)C\gamma_5d_b(x)]u_c(x),$$
where $u$ and $d$ are the quark operators, $a,b,c$ are the color indices, $T$ denotes
a transpose and $C$ is the charge conjugation matrix, while $t$ is an arbitrary parameter.

Following \cite{2}, we shall use the current determined by Eq.(7) with $t=-1$.  It can be written
(up to a factor $1/2$ ) as \cite{2}
\begin {equation}
j(x)= \varepsilon_{abc}[u_a^T(x)C\gamma_{\mu}u_b(x)]\gamma_5\gamma^{\mu}d_c(x)
\end{equation}
This choice was shown in \cite{BL} to be most relevant for
description of nucleons, since the polarization operator
$\Pi(q^2)$ calculated with this current satisfies two main
requirements. On the RHS of Eq.(5) the contribution of the nucleon
pole exceeds that of the higher states (approximated by
continuum). On the LHS of Eq.(5) the higher order terms of
$M^{-2}$ series should drop fast enough to be consistent with the
convergence of the OPE.  It is important also that there is a gap
between the position of the lowest pole $m^2$ and the effective
threshold $W^2$. This choice of the current was advocated in
\cite{6a}. There are many papers, in which this very current was
used - see, e.g., \cite{R}.

Several lowest terms of the OPE for the current (8) have been calculated in \cite{2,2c}. The
leading term depends on $q^2$ as $q^4\ln q^2$. It comes from the free
three-quark loop. The higher order OPE terms contain the matrix
elements
$$
\langle0|\bar qq|0\rangle\,, \quad \langle0|\frac{\alpha_s}\pi\,
G^a_{\mu\nu}G^a_{\mu\nu}|0\rangle\,, \quad \langle0|\bar qq\bar
qq|0\rangle\,,
$$
etc. Analysis carried out in \cite{2c} contained also the most important
radiative corrections in which the QCD coupling constant $\alpha_s$ is
multiplied by ``large logarithm" $\ln q^2$. Corrections of the order
$(\alpha_s\ln q^2)^n$ to the leading OPE terms have been calculated earlier  in
\cite{8}.

 The appropriate interval
\begin{equation}
0.8\mbox{ GeV}^2\ <\ M^2\ <\ 1.4\mbox{ GeV}^2
\end{equation}
("Borel window") was found in \cite{2c}. The values of the lowest OPE terms at conventional values of
the QCD condensates enable to expect the convergence of OPE series. Also, for the solution
found in \cite{2c} the contribution of the pole exceeds that of the continuum.

The OPE series for $\Pi^q$ and $\Pi^I$
start from the terms $q^4\ln q^2$ and $q^2\ln q^2$ correspondingly. Thus, in somewhat straightforward
interpretation of OPE only $\Pi^q$ contains a leading term. This allows
to consider the chirality conserving structure $\Pi^q$ as a more important one.
However actually we consider the values of the Borel mass $M^2$ of the order of the proton mass.
After the Borel transform the leading contributions to $\Pi^q$ and $\Pi^I/m$ are of the
same order. The chirality violating sum rule for $\Pi^I$ can be considered as important
as that for $\Pi^q$. The two sum rules were considered on  the same terms, requiring the same
accuracy for both of them. That's why the terms corresponding to $\Pi^q$ and $\Pi^I$
on the right hand side of Eq.(6) were included with the same weights.


A weak point of this procedure is that the choice of parameters
$m,\lambda^2$ and $W^2$ may be not simple.
There can be several local minima of $\chi^2$ corresponding to several
sets of the parameters.

Note that the values of QCD condensates are known with large
uncertainties. The expectation value $\langle0|\bar qq|0\rangle$
can be determined with the larger accuracy than the other
condensates due to the Gell-Mann--Oakes--Renner relation~\cite{9}.
There is no experimental data on values of four- and six-quark
condensates. They can be calculated in the factorization
approximation for the case of large number of colors $N_c\gg1$.
The accuracy of this approximation for $N_c=3$ is obscure. The
value of gluon condensate was initially obtained from the QCD sum
rules for $\rho$ mesons \cite{10}. However the sum rules for other
mesons lead to somewhat smaller \cite{11} or larger \cite{12}
values, with the latest analysis presented in \cite {BL1}. Hence
it is reasonable to study dependence of the nucleon parameters on
the values of the QCD condensates.

This dependence should be investigated together with inclusion of
radiative corrections. The latter imitate modification of the values of
the condensates providing the contributions $\sim\alpha_s$ and
$\alpha_s\ln q^2$ to each of OPE term. Thus, including the radiative
corrections into our analysis, we can separate the effects of
uncertainties in the QCD condensates values.

In the present paper we analyze the role of the unphysical
solution for the nucleon QCD sum rules, which corresponds to the
continuum contribution exceeding that of the pole. We mentioned
this solution in our earlier papers \cite{13,14}. Here we show
that both physical and unphysical solutions provide local minima
of the function $\chi^2(m, \lambda_N^2, W^2)$ determined by
Eq.~(6). For the absolute values of condensates, which differ
noticeably from the conventional values (still consistent with
convergence of the OPE series on the LHS
 of the sum rules), the minima corresponding to physical solutions
may vanish.

We show that inclusion of the radiative corrections modifies the
situation. One could expect this, since it was shown in \cite{13}
that the radiative corrections affect mostly the value of the
nucleon residue $\lambda_N^2$. After the corrections of the order
$\alpha_s$ are included perturbatively, the physical solution
exists for a broader interval of the values of condensates. Also,
domination of continuum contribution over that of the pole for the
unphysical solution becomes stronger. The unphysical solution
becomes "more unphysical".

Note that the problem of the pole dominance emerged in other QCD sum rules studies.
In review on the QCD sum rules analysis of the pentaquark states
\cite{BR1} the authors found that it is very difficult to satisfy the requirements
of pole domination, OPE series convergence and stability within the Borel window simultaneously. This contrasts the earlier statements (cited in \cite {BR1}) that the QCD sum rules support the existence of the pentaquark.
However, the authors of \cite{BR1} do not make a definite statement on
the QCD sum rules predictions about the pentaquark states.
In the QCD sum rules analysis of the light tetraquark states \cite{BR2} it was found that the solution with the domination of the pole can be obtained only for small values of the Borel mass $M^2$,
where the OPE series does not converge.
The authors of \cite{BR2} conclude the QCD sum rules analysis does not support existence of light tetraquark particles.
In view of the analysis carried out in the present paper, inclusion of the radiative corrections
may become important here.

In Section 2 we analyze the interplay of the physical and
unphysical solutions taking into account only the leading
radiative corrections. We include corrections of the order
$\alpha_s$ in Sec.~3. We summarize in Sec.~4.

\section{Interplay of the physical and unphysical solutions}

We start by representing the nucleon QCD sum rules \cite{2c}
without using the factorization hypothesis for the condensates of
the high dimensions.
 The LHS of the Borel transformed nucleon sum rules (Eq.(5)) with inclusion
of the anomalous dimensions (i.e. of the corrections of the order
($\alpha_s\ln q^2)^n$) can be written as
\begin{equation}
{\cal L}^q=\sum_n\tilde A_n(M^2)\,, \quad {\cal L}^I=\sum_n\tilde
B_n(M^2).
\end{equation}
Here the lower indices show the dimensions of the condensates. If
the current $j$ in Eq.(1) is given by Eq.(8), the terms on the
right hand sides of Eq.~(10) are \cite{2, 2c}
\begin{eqnarray}
&& \tilde A_0=\frac{M^6E_2}L\,, \quad \tilde A_4=\frac{c_4M^2E_0}{4L}\,,
\quad \tilde A_6=\frac43\,c_6L\,,\quad
\tilde A_8=-\frac13\,\frac{c_8}{M^2}\,,
\nonumber\\
&&  \tilde B_3=2c_3M^4E_1\,, \quad \tilde B_7=-\frac{c_7}{12}\,, \quad
\tilde B_9=\frac{272}{81}\,\frac{c_9}{M^2}\,.
\end{eqnarray}
Here $L$ accounts for the leading radiative corrections
$\sim\alpha_s\ln q^2$ \cite{8}
$$
L\ =\ \left(\frac{\ln q^2/\Lambda^2_{\rm
QCD}}{\ln\mu^2/\Lambda^2_{\rm QCD}}\right)^\gamma,
$$
with the anomalous dimension $\gamma=4/9$ ($L=1$ if the leading
radiative corrections are neglected). After the Borel transform
\begin{equation}
L\ =\ \left(\frac{\ln M^2/\Lambda^2_{\rm QCD}}{\ln\mu^2/\Lambda^2_{\rm
QCD}}\right)^{4/9}.
\end{equation}

The condensates $c_i$ are
\begin{eqnarray}
&& c_3=-(2\pi)^2\,\langle0|\bar qq|0\rangle\,, \qquad
c_4=(2\pi)^2\Big\langle0\Big|\frac{\alpha_s}\pi\,G^2\Big|0\Big\rangle,
\nonumber\\
&& c_6=(2\pi)^4\langle0|\bar qq\bar qq|0\rangle\,, \qquad
c_7=-(2\pi)^4\,\Big\langle0\Big|\bar q\frac{\alpha_s}\pi
G^2q|0\Big\rangle\,,
\nonumber\\
&& c_8=(2\pi)^4 \Big\langle0\Big|\bar
qq\bar q\frac{\alpha_s}\pi\,
G^a_{\mu\nu}\,\frac{\lambda^a}2\,\sigma_{\mu\nu}q\Big|0\Big\rangle\,,
\quad c_9=-(2\pi)^6\frac{\alpha_s}\pi\,\langle0|\bar qq\bar qq \bar qq
|0\rangle\,.
\end{eqnarray}
Note that the structure of the condensates $c_6$ and $c_9$ is
indeed more complicated. For example, $c_6$ contains the
condensates $\langle0|\bar q\Gamma^Aq\bar q\Gamma^Aq|0\rangle$
with $\Gamma^A$ being the basic $4\times4$ matrices $(\Gamma^A=I,
\gamma_\mu,\gamma_5, i\gamma_5\gamma_\mu, \sigma_{\mu\nu})$. The
same is true for $c_9$. Hence the matrix elements $\langle0|\bar
qq\bar qq |0\rangle$ and $\langle0|\bar qq\bar qq\bar qq|0\rangle$
in expressions for $c_6$ and $c_9$ in Eq.~(13) are somewhat
``effective" $4q$ and $6q$ condensates. Denote
\begin{eqnarray}
&& c_3=a_0f_{2q}\,, \quad c_4=b_0f_b\,, \quad c_6=a^2_0f_{4q}\,,
\nonumber\\
&& c_7=a_0b_0f_{qg}\,, \qquad c_9=\frac{\alpha_s}\pi\,a^3_0f_{6q}\,.
\end{eqnarray}
Here we introduced dimensionless parameters $f_i$. In the factorization
approximation $f_{4q}=f^2_{2q}$, $f_{qg}=f_{2q}f_b$, $f_{6q}=f^3_{2q}$.
We shall investigate the dependence of nucleon parameters on QCD
condensates modifying the values of $f_i$. Note also that
$c_8=\mu^2_0a^2_0$ with $\mu^2_0\approx 0.8\,\rm GeV^2$ \cite{2}.

Following \cite{2} we write
\begin{equation}
a=-(2\pi)^2\langle0|\bar qq|0\rangle\,, \qquad
b=(2\pi)^2\Big\langle0\Big|\frac{\alpha_s}\pi\,G^2\Big|0\Big\rangle\,.
\end{equation}
For traditional choice of the normalization point $\mu=0.5$~GeV
\begin{equation}
a=a_0=0.55\,{\rm GeV}^3\,, \qquad b=b_0=0.5\rm\,GeV^4\,.
\end{equation}
Putting all $f_i=1$ in Eq.~(14) we come to the standard nucleon sum
rules \cite{2,2c} with
\begin{eqnarray}
&& \tilde A_0=\frac{M^6E_2}L\,, \quad \tilde A_4=\frac{bM^2E_0}{4L}\,,
\quad \tilde A_6=\frac43\,a^2L\,, \quad \tilde
A_8=-\frac13\frac{\mu^2_0}{M^2}\,a^2\,,
\nonumber\\
&& \tilde B_3=2aM^4E_1\,, \quad\tilde B_7=-\frac{ab}{12}\,, \quad
\tilde B_9=\frac{272}{81}\frac{\alpha_s}\pi\frac{a^3}{M^2}\,.
\end{eqnarray}
Here
$$
E_n=E_n(x)=1-e^{-x}\sum^n_{k=0} \frac{x^n}{n!}\,, \quad
n=0,1,2\,, \quad x=W^2\Big/M^2\,,
$$
i.e. the contributions of continuum are transferred to the LHS of
Eq.~(5). The RHS of Eq.~(5)
\begin{equation}
R^q(M^2)=\lambda^2\,e^{-m^2/M^2}\,, \qquad
R^I(M^2)=m\lambda^2e^{-m^2/M^2}
\end{equation}
contain only the contribution of the nucleon pole with
$\lambda^2=32\,\pi^4\lambda^2_N$.

In the Borel window
\begin{equation}
\frac{|\tilde A_8|}{|\tilde A_6|} \approx \frac{1}{4}; \quad \frac{\tilde B_9}{\tilde B'_3} \ll 1,
\end{equation}
with $\tilde B'_3=2aM^4$ is just $\tilde B_3$ determined by Eq.(17) before the threshold
contribution is transferred to the LHS. (Note that $\tilde A_4$ and $\tilde B_7$ are numerically small
due to a small coefficient connected with contributions of the gluon condensate). This is consistent with
the hypothesis about convergence of the OPE series.

We can expect that the convergence will not be spoiled
by the disregarder terms. Note that the term $A_8$ can be viewed as coming from expansion of the
expectation value of the operator $\bar q(0)q(x)$ in powers of $x^2$. The ratio $A_8/A_6$ is the characteristic scale
for further expansion in powers of $M^{-2}$, caused by expansion in powers of $x^2$. The mixed quark-gluon condensates
of higher dimensions are expected to have small numerical factors, connected with the gluons, similar to
$c_4$ and $c_7$. Finally, there is the only QCD parameter $\Lambda_{QCD}$, and there can be contributions
containing the factor $\Lambda_{QCD}^2/M^2 \ll 1$.

In framework of the factorization hypothesis the method developed
in \cite{2,2c} provides
\begin{equation}
m=0.931\mbox{ GeV}\,, \quad \lambda^2=1.86\mbox{ GeV}^6, \quad
W^2=2.09\mbox{ GeV}^2\,,
\end{equation}
if the numerical values (16) are employed. We assume $\Lambda_{\rm
QCD}\approx150\,$MeV. In the one-loop approximation this
corresponds to $\alpha_s(1$~GeV$^2)\approx 0.37$, which is
consistent with the PDG data \cite{RMF}.

If the radiative corrections are neglected, i.e. $L=1$ the
solution appears to be
\begin{equation} m=0.930\mbox{ GeV}\,, \quad \lambda^2=1.79\mbox{
GeV}^6\,, \quad W^2=2.00\mbox{ GeV}^2\,.  \end{equation} One can
see that for the solution represented by Eq.~(21) the contribution
of the pole exceeds that of continuum more than twice.

However the successive inclusion of the OPE terms leads to another
solution. Taking into account only the condensates with the dimensions
$d=3,\,4$ we find a trivial solution $m=\lambda^2=W^2=0$. Inclusion of
the condensate with $d=6$ keeps $m=0$, $W^2=0$ but provides
$\lambda^2=4/3\,a^2=0.4\rm\,GeV^6$. The condensates with $d=7$ and
$d=8$ require the nonzero values of $m$ and $W^2$, i.e.
\begin{equation}
m=0.6\mbox{ GeV}\,, \quad \lambda^2=0.79\mbox{ GeV}^6\,, \quad
W^2=1.0\mbox{ GeV}^2\,.
\end{equation}
We treat this solution as an unphysical one since in the Borel
window determined by Eq.(9) the contribution of the pole is
smaller than that of the continuum. For example, at the
characteristic value $M^2=1~$GeV$^2$ of the Borel window the ratio
of the continuum and pole contributions is 2 and 1.75 for the
$\Pi^q$ and $\Pi^I$ structures correspondingly.

As we shall see below, the unphysical solution manifests itself
when the QCD condensates deviate from their conventional values.
Note that for all the cases discussed below Eq.(19) is still true,
and thus the convergence of the OPE series is not violated. Also
(see, e.g. Fig.1 below) the value of $\chi^2$ is small enough.
Hence, the only reason for assuming this solution to be unphysical
is the domination of the contribution of the continuum over that
of the pole.

It is instructive to trace the dependence of nucleon parameters on
the value of gluon condensate. At $f_b=1$, i.e. at the value of
the gluon condensate corresponding to Eq.~(16) the functional (6)
has two local minima corresponding to the solutions (21) and (22).
The deeper minimum is provided by the unphysical solution (22). At
$f_b<1$ we still have two minima, and the values of the nucleon
parameters do not change much even for $f_b=0$ due to a relatively
small value of the term $\tilde A_4$ in Eq.~(17). However for
$f_b>0.2$ the deeper minimum corresponds to the unphysical
solution (22). Somewhat straightforward employing of the
chi-squired method may leave the physical solution unnoticed. The
situation is more dramatic for $f_b>1$. At $f_b>1.04$ the certain
minimum corresponding to the physical solution vanishes and only
the unphysical solution survives.

In order to illustrate the role of the unphysical solution in the
Borel window we introduce the function
$$m(M^2)=\frac{{\cal L}^I(M^2)}{{\cal L}^q(M^2)},$$
with ${\cal L}^I$ and ${\cal L}^q$ defined by Eq.(10).  Using
Eq.(18) we see that for the values of $M^2$ determined by Eq.(9)
we can expect $m(M^2) \approx const=m$ for the solutions of the
sum rules equations. Putting $f_b=0.6$ (to make the difference of
the corresponding $\chi^2$ values more visible) we expect the
solutions to be close to those described by Eqs.(21),(22) for
$f_b=1$. The functions $m(M^2)$ for $W^2=2.09$~GeV$^2$- see
Eq.(21) and for $W^2=1.0~$GeV$^2$ -see Eq.(22) are shown in Fig.3.
One can see that the unphysical solution exhibits a more stable
behavior then a physical one.

In Fig. 3 we show the dependence of $m,\lambda^2$ and $W^2$ on the
value of $f_b$.

Note that if there is an unphysical solution which corresponds to
the absolute minimum of the functional $\chi^2(m, \lambda^2, W^2)$
defined by Eq.(6), the search for another solution and its
interpretation becomes more complicated. It is rather simple if
both minima are sharp. However, if the unphysical solution
corresponds to a wide minimum, while the second solution
corresponds to a shallow one, interpretation of the latter
solution requires additional analysis.

Now we come to variation of the values of the quark condensates. Let us
first modify the value of the condensate $\langle0|\bar qq|0\rangle$,
keeping the other ones to be unchanged. Hence we put
$f_{4q}=f_{6q}=f_b=f_{qg}=1$ in Eq.~(14), changing the value of
$f_{2q}$. The dependence of the nucleon mass on $f_{2q}$, somewhat
``partial derivative" with respect to $\langle0|\bar qq|0\rangle$ is
shown in Fig.~4. The physical solution exist only if $f_{2q}>0.99$.

We investigate dependence on the parameters $f_{4q}$ and $f_{6q}$
in the same way. The physical solution disappears if $f_{4q}$
exceeds slightly the value $f_{4q}=1$ corresponding to the
factorization hypothesis. There is no minimum corresponding to a
physical solution  for $f_{4q}>1.01$ -- Fig.~5. On the contrary,
the physical solution is not consistent with small values of
$f_{6q}$. It requires $f_{6q}>0.96$ -- Fig.~6.

It is instructive also to investigate the dependence of the
nucleon mass on the quark scalar condensate $\langle0|\bar
qq|0\rangle$ assuming the factorization hypothesis. In this case
$f_{4q}=f^2_{2q}$, $f_{qg}=f_{2q}$ and $f_{6q}=f^2_{3q}$ while
$f_b=1$ in Eq.~(14). The result is shown in Fig.~7. There is no
minimum corresponding to a physical solution since it vanishes for
$f_{2q}>1.35$.

\section{Inclusion of the radiative corrections of the order
\boldmath$\alpha_s$}

Here we include the corrections of the order $\alpha_s$ and
$\alpha_s\ln q^2$ in the lowest order of perturbation theory. This
modifies the contributions to $\Pi^i(q^2)$. For the contributions
of the free quark loop $A_0$ and for those of the scalar quark
condensate $B_3$ and the four-quark condensate $A_6$ we have now
\cite{15,16}
\begin{eqnarray}
A_0 &=& -\frac1{64\pi^4}Q^4\ln\frac{Q^2}{\mu^2}\left(1+\frac{71}{12}
\frac{\alpha_s}\pi-\frac12 \frac{\alpha_s}\pi\ln \frac{Q^2}{\mu^2}
\right),
\nonumber\\
A_6 &=& \frac23\,\frac{\langle0|\bar qq|0\rangle^2}{Q^2} \left(1-
\frac56\,\frac{\alpha_s}\pi -\frac13\,\frac{\alpha_s}\pi
\ln\frac{Q^2}{\mu^2}\right),
\\
B_3 &=& -\frac{\langle0|\bar qq|0\rangle}{4\pi^2}\,Q^2\ln
\frac{Q^2}{\mu^2}\left(1+\frac32\,\frac{\alpha_s}\pi\right),
\end{eqnarray}
with $Q^2=-q^2$. Corrections to the other OPE terms are not included
because of the large uncertainties of the values of the condensate.

The numerically large coefficient of the correction of the order
$\alpha_s$ to the term $A_0$ caused doubts in convergence of OPE
series \cite{17}. In \cite{16} Eqs. (23) and (24) were used for
determination of nucleon parameters in framework of the finite
energy sum rules. Inclusion of the radiative correction was shown
to diminish the value of the nucleon mass, assuming that the
threshold value $W^2$ does not change. Similar result was obtained
in \cite{18} in framework of the Borel transformed sum rules.
However the authors of \cite{13} demonstrated that the radiative
corrections modify mostly the values of $\lambda^2$ and $W^2$,
without important influence on the value of nucleon mass.

For the contributions to the Borel transformed sum rules we can write
\cite{13}
\begin{eqnarray}
&& \hspace*{-1cm}
\tilde A_0(M^2,W^2)\ =\ M^6E_2\left[1+\frac{\alpha_s}\pi
\left(\frac{53}{13}-\ln\frac{W^2}{\mu^2}\right)\right]
\nonumber\\
&& -\ \frac{\alpha_s}\pi \left[M^4W^2
\left(1+\frac{3W^2}{4M^2}\right)e^{-W^2/M^2}+M^6{\cal E}
\left(-\frac{W^2}{M^2}\right)\right],
\nonumber\\
&& \hspace*{-1cm}
\tilde A_6(M^2,W^2)\ =\ \frac43a^2\left[1-\frac{\alpha_s}\pi
\left(\frac56+\frac13\left(\ln\frac{W^2}{\mu^2}+{\cal E}
\left(-\frac{W^2}{\mu^2}\right)\right)\right)\right],
\nonumber\\
&& \hspace*{-1cm}
\tilde B_3(M^2,W^2)\ =\ 2aM^4E_1\left(1+\frac32\,\frac{\alpha_s}\pi
\right),
\end{eqnarray}
with ${\cal E}(x) = \sum\limits_{n=1} \frac{x^n}{n\cdot n!}\,$.

It was shown in \cite{13} that the results do not change much if
we put $\alpha_s(M^2)=\alpha_s(1\rm\,$GeV$^2)$. As we said above,
we assume $\alpha_s(1 $GeV$^2)=0.37$, which is consistent with the
recent data presented in Fig.9.2 of \cite{RMF}. A somewhat larger
value is given in \cite{BL1}. Anyway, the nucleon parameteres
depend weakly on the actual value of $\alpha_s$-see Fig.2 of
\cite{13}. Hence the same referes to the present analysis.

The physical solution is now
\begin{equation}
m=0.94\mbox{ GeV}, \quad \lambda^2=2.00\mbox{ GeV}^6, \quad
W^2=1.90\mbox{ GeV}^2\,.
\end{equation}
The unphysical solution is
$$m=0.60 ~GeV, \quad \lambda^2=0.56~ GeV^6,\quad
W^2=0.70~ GeV^2,$$ with the contribution of the pole 3 times
smaller than that of the continuum. Hence the radiative
corrections made the unphysical solution even more "unphysical".
 Dependence of the nucleon mass on the values of
condensates is shown in Figs.~2a, 4--7. One can see that the
physical solution exists in the larger interval of the values of
$f_i$ than in the case when only anomalous dimensions are
included. The physical solution (described by Eq.(26) for
$f_b=f_{4q}=f_{6q}=1$) provides absolute minimum of the function
$\chi^2$ for $f_b<1.8$, $f_{4q}<1.35$ and $f_{6q}>0.2$ -- see
Figs.~2, 5--7. Dependence on the value $f_{2q}$ without and with
factorization assumptions is shown in Figs. 4 and 7
correspondingly.

Note also that the variation of the limits of the Borel window
changes the values of nucleon parameters only by  several percent.
For example, solving Eq.(5) in the intervals
$0.7~$GeV$^2<M^2<1,8$GeV~$^2$ we find $m=0.946$GeV, while in the
interval $1.0~$GeV$^2<M^2<1,6$GeV~$^2$ we obtain $m=0.953$ GeV for
$f_b=f_{2q}=f_{4q}=1$. There is also an unphysical solution, but
the corresponding values of $\chi^2$ are at least ten times larger
then for the physical one.

\section{Summary}

We analyzed dependence of the nucleon mass on the values of QCD
condensates in framework of the QCD sum rules. We investigated also the
dependence of the residue of the nucleon pole and of the continuum
threshold in ``pole + continuum" model of the hadron spectrum. These
dependences were studied with inclusion of radiative corrections of the
order $\alpha_s$ and $\alpha_s\ln q^2$.

We presented three sets of the results for minimization of $\chi^2$ defined by Eq.(6)
in the Borel window defined by Eq.(9).
They correspond to total
neglection of the radiative corrections, to inclusion of the
corrections $(\alpha_s\ln q^2)^n$ in all orders and to
perturbative inclusion of corrections $\sim\alpha_s$ and
$\alpha_s\ln q^2$.

It is shown that even at relatively small deviations of QCD condensates
from the standard values the QCD sum rules have an unphysical solution
with the contribution of continuum exceeding several times that of the
nucleon pole. This contradicts the idea of the ``pole + continuum"
model for the hadron spectrum.

We showed that (neglecting the radiative corrections) at $f_b<1$
there is an interplay of the physical and unphysical solutions. In
this case the two solutions can be separated. If both unphysical
and physical solutions are connected with sharp minima of the
dependence of the function $\chi^2$  on nucleon parameters, they can be
separated easily. However in the general case, for example, for a
wide minimum corresponding to the unphysical solution or for a
shallow minimum of another solution the interpretation of the
latter one may be obscure.

The strongest limits on the values of condensates corresponding to
a physical solution emerge if one includes the radiative
corrections $(\alpha_s\ln q^2)^n$, the limits are weaker if the
radiative corrections are totally neglected and still weaker if
the corrections are included perturbatively -- Figs.~1,3--7. Thus
for consistent calculations it is reasonable either to include the
radiative corrections perturbatively or just to ignore them.

We demonstrated that for the physical solution the value of the nucleon
mass is less sensitive to the exact values of the condensates than
those of the residue $\lambda^2$ and of the continuum threshold $W^2$.
In other words, the uncertainties in the value of condensates influence
mostly the magnitudes of $\lambda^2$ and $W^2$.

We found that the nucleon mass depends mostly on the expectation value of the
scalar quark operator $\bar qq$. Dependence on the condensates of
higher dimensions is much weaker. The dependence on the condensate
$\langle0|\bar qq|0\rangle$ becomes weaker if one assumes
factorization hypothesis for the four-quark and six-quark
condensates.

In the QCD sum rules  the nucleon mass obtained a nonzero value
due to exchange by noninteracting quark--antiquark pairs between
the nucleon current and vacuum. As we have seen perturbative
inclusion of interactions taking place at the distances of the
order $1/M\sim1\rm\,GeV^{-1}$, corresponding to inclusion of the
radiative corrections $\sim\alpha_s$ make the solution more
stable. The physical solution exists in a broad interval of the
values of $\langle0|\bar qq|0\rangle$ if the radiative corrections
are included perturbatively. On the other hand the unphysical
solution becomes "more unphysical", with a stronger dominance of
the contribution of continuum over that of the pole. This may be
important for the QCD sum rules analysis of the many-quark systems
carried out nowadays \cite{BR1,BR2}.

We thank B. L. Ioffe for useful comments. We acknowledge the
partial support by the RSGSS grant~3628.2008.2.

\newpage
\section{Figure captions}
\noindent Fig.1 Dependence of the nucleon mass and of $\chi^2$ per
degree of freedom (assuming 1\% error bar)on value of the gluon
condensate $f_b$. The solid lines correspond to the physical
solution. The dashed lines correspond to the unphysical solution.
For convenience the values of $\chi^2$ are reduced by the factor
of 5,
 $m_0=0.93~$GeV$^2$ is the value of $m$ corresponding to physical
 solution for $f_b=1$.
 For $f_b>1.04$ only the latter one exists.

 \noindent
Fig.2
Behavior of the function $m(M^2)$ for the physical (solid
line) and unphysical (dashed line) in the case $f_b=0.6$.

 \noindent
 Fig.3
 Dependence of the nucleon parameters on gluon
condensate. Figs. a,b,c show the mass $m$, residue $\lambda^2$ and
threshold $W^2$ correspondingly. Dashed curves show the case with
all the radiative corrections neglected, dotted curves correspond
to inclusion of the anomalous dimensions. Solid curves are for
perturbative inclusion of the corrections $\sim\alpha_s$. The
horizontal dashes denote the transitions from the physical
solutions to unphysical ones.

\noindent
 Fig.4
 Dependence of the nucleon mass on the parameter
$f_{2q}$. Only the term $\tilde B_3$ in Eqs. (11) and (25) is
modified. The notations are the same as in Fig.~3.

\noindent
 Fig.5
 Dependence of the nucleon mass on the parameter
$f_{4q}$. Only the term $\tilde A_6$ in Eqs. (11) and (25) is
modified. The notation are the same as in Fig.~3.

\noindent
 Fig.6
 Dependence of the nucleon mass on the parameter
$f_{6q}$. Only the term $\tilde B_9$ in Eq.~(11) is modified. The
notations are the same as in Fig.~3.

\noindent
Fig.7
 Dependence of the nucleon mass on the parameter $f_{2q}$
under the factorization hypothesis $f_{4q}=f^2_{2q}$,
$f_{6q}=f^3_{2q}$. The notations are the same as in Fig.~3.

\newpage

\begin{figure}[h]
\centerline{\epsfig{file=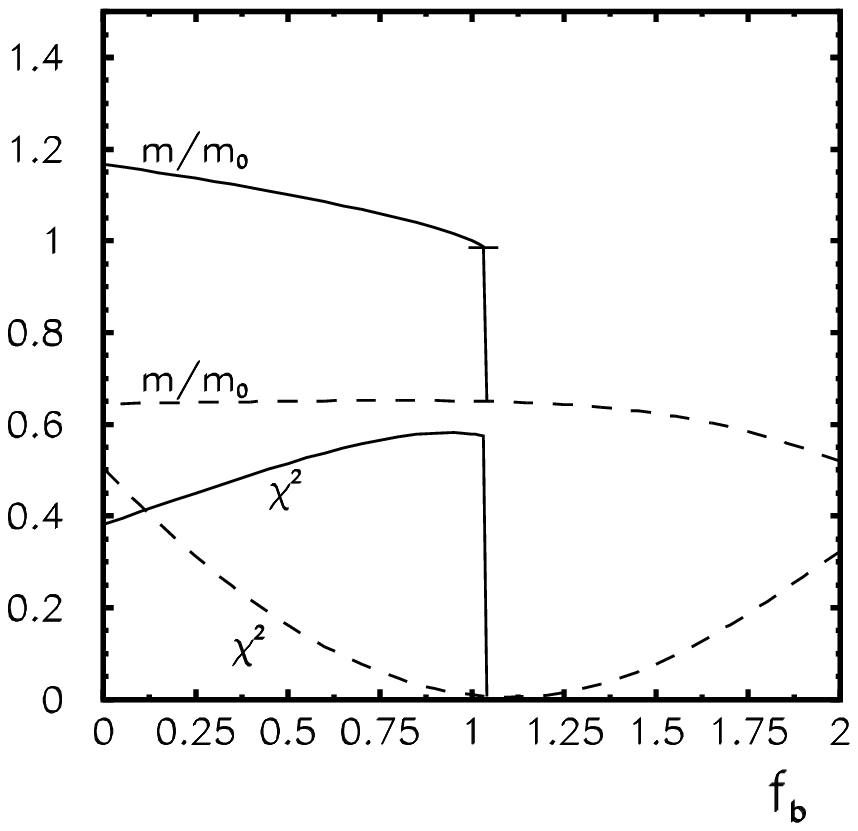,width=7cm}}
 \caption{}
 \end{figure}

\begin{figure}[h]
\centerline{\epsfig{file=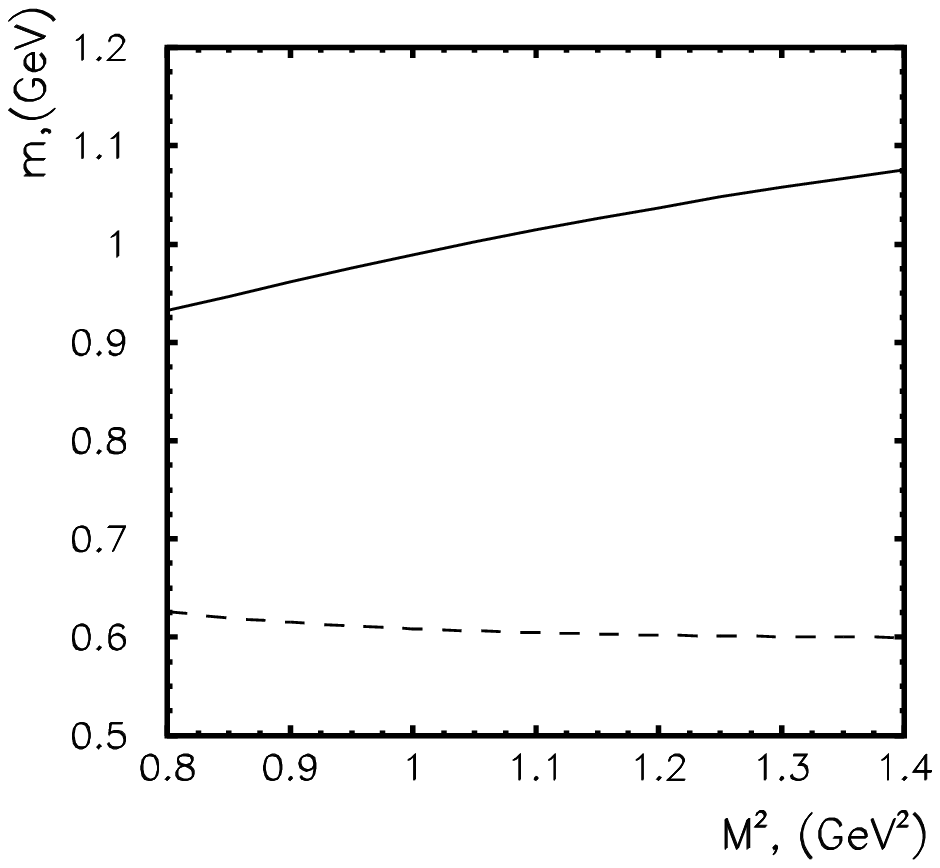,width=7cm}}
 \caption{}
 \end{figure}

\begin{figure}[h]
\centerline{\epsfig{file=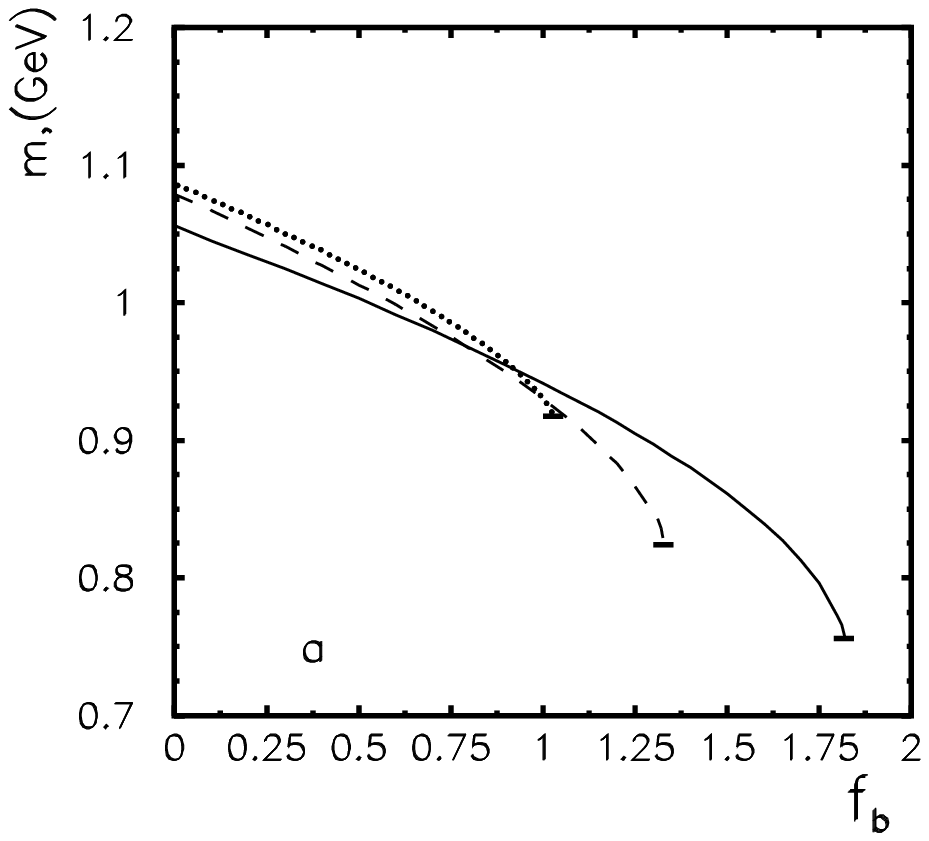,width=7cm}
\epsfig{file=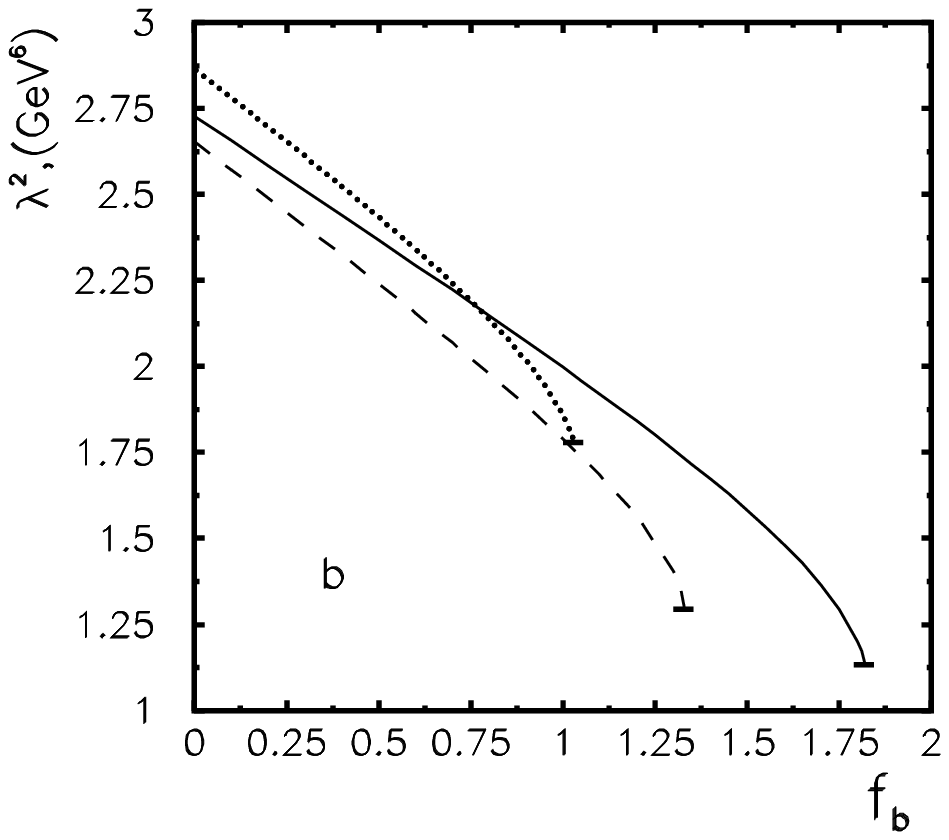,width=7cm}}
\centerline{\epsfig{file=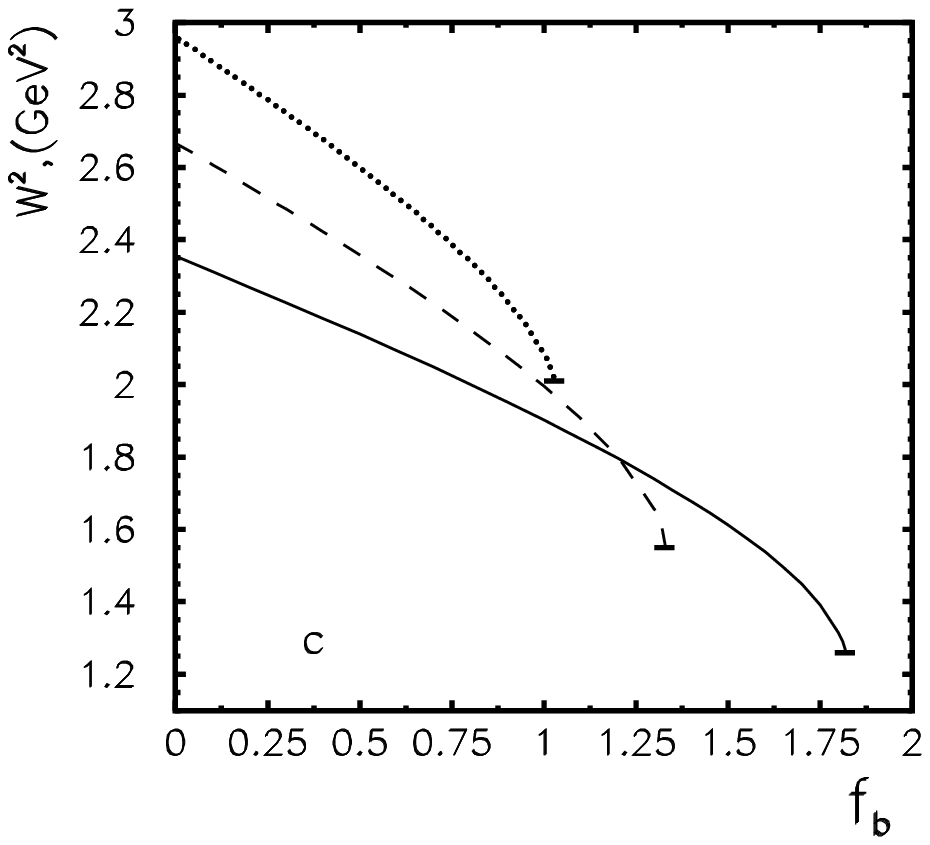,width=7cm}}
 \caption{}
\end{figure}

\begin{figure}[h]
\centerline{\epsfig{file=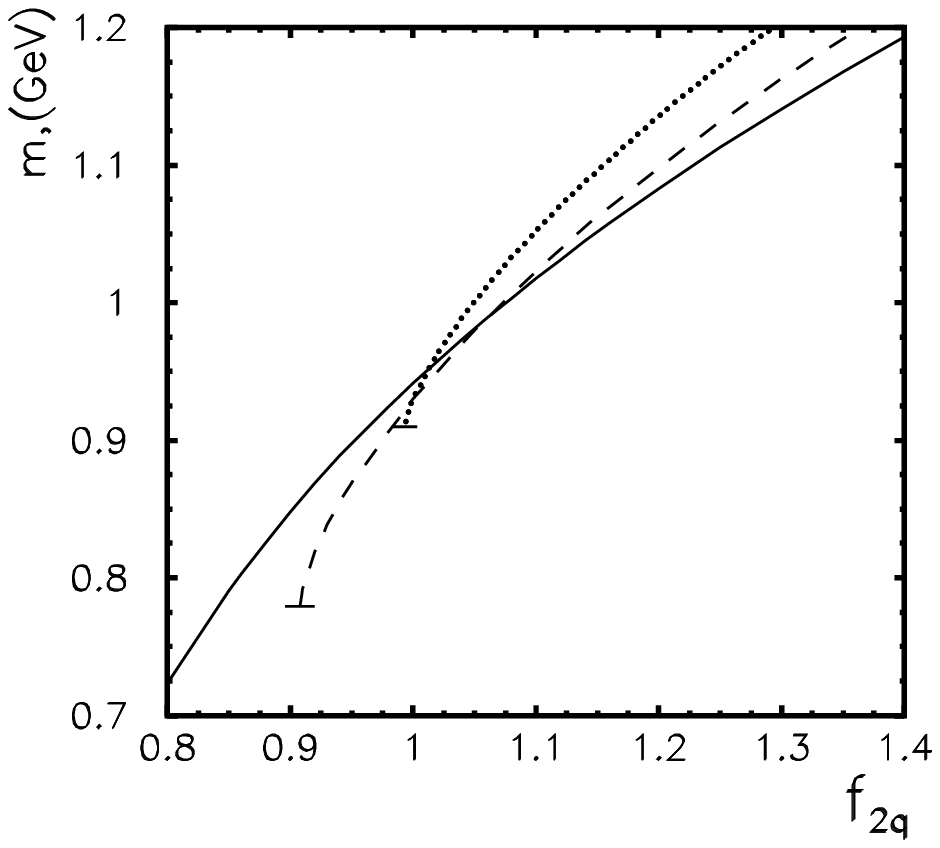,width=7cm}}
 \caption{}
\end{figure}

\begin{figure}[h]
\centerline{\epsfig{file=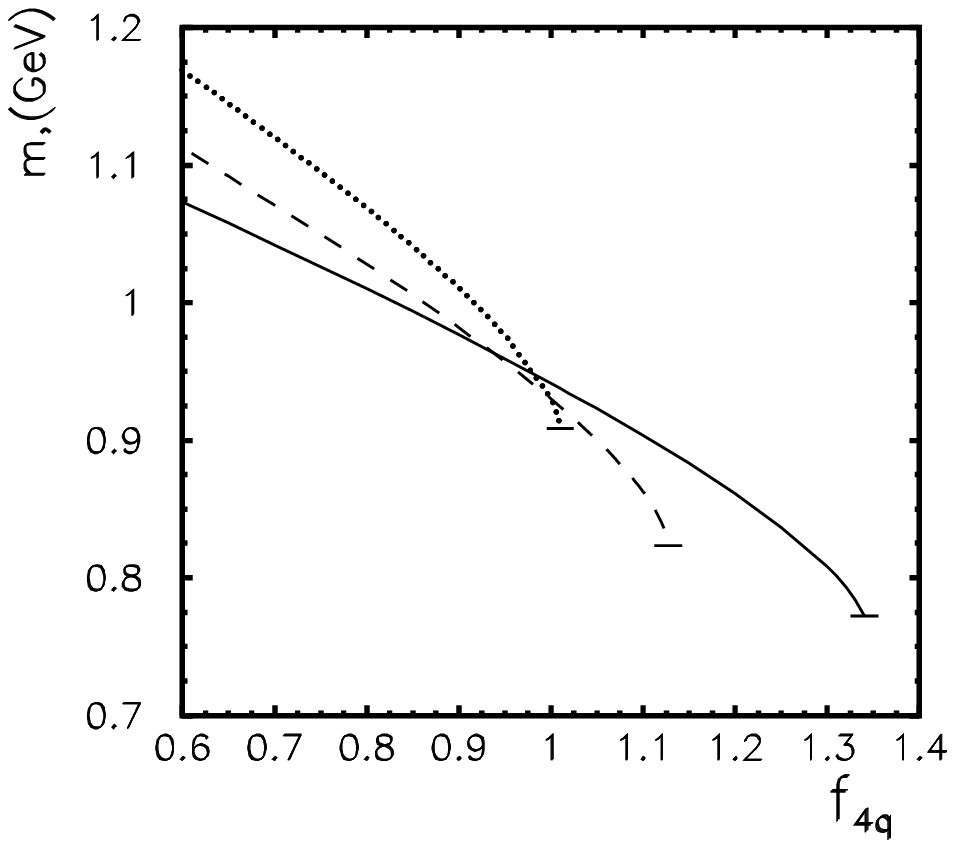,width=7cm}}
  \caption{}
\end{figure}

\begin{figure}[h]
\centerline{\epsfig{file=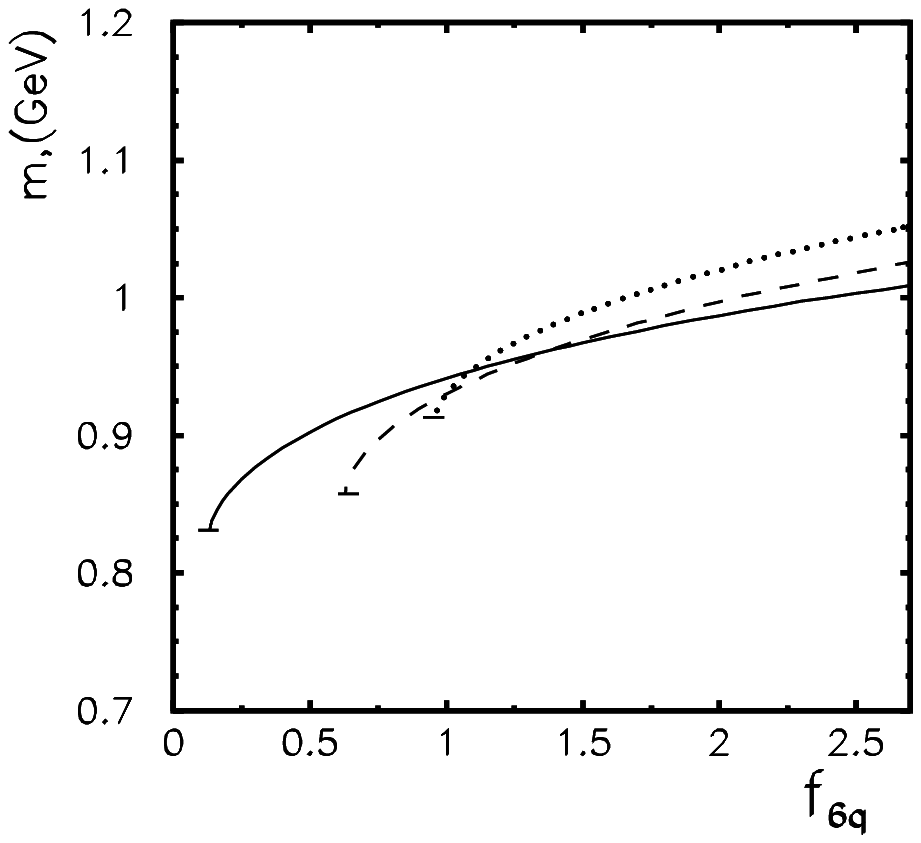,width=7cm}}
  \caption{}
\end{figure}

\begin{figure}[h]
\centerline{\epsfig{file=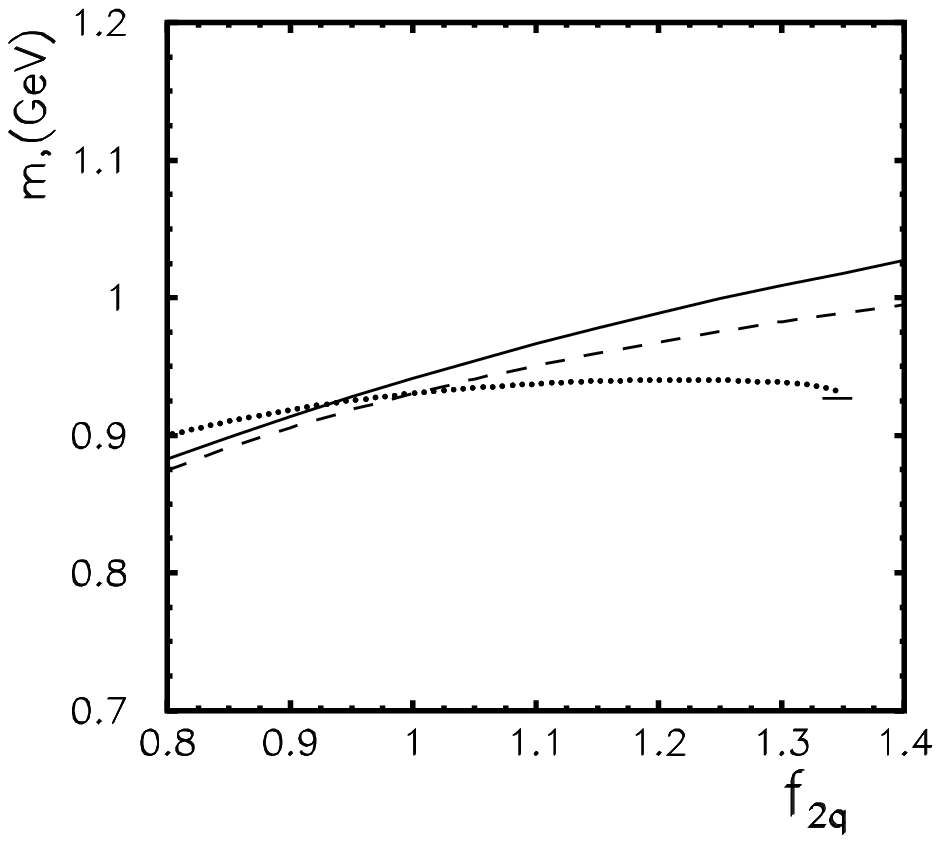,width=7cm}}
 \caption{}
\end{figure}


\newpage
\begin{thebibliography}{**}

\bibitem{1} M.~A.~Shifman, A.~I.~Vainshtein, and V.~I.~Zakharov, Nucl.
Phys. B~{\bf147}, 385 (1979).

\bibitem{2} B.~L.~Ioffe, Nucl. Phys.  B~{\bf188}, 317 (1981); E {\bf191},
591 (1981).

\bibitem{2a} Y.~Chung, H.~G.~Dosch, M.~Kremer and D.~Schall, Phys. Lett. B {\bf 102},
175 (1981).

\bibitem{2b} Y.~Chung, H.~G.~Dosch, M.~Kremer and D.~Schall, Nucl. Phys. B {\bf 197},
55 (1982).

\bibitem{2c} B.~L.~Ioffe and A.~V.~Smilga, Nucl. Phys. B~{\bf232}, 109
(1984).

\bibitem{2d} H.~G.~Dosch, M.~Jamin and S.~Narison, Phys. Lett B~{\bf220}, 251
(1989).

\bibitem{3a} E.~Bagan, M.~Chabab, H~G.~Dosch and S.~Narison, Phys. Lett B.~{\bf 301}, 243 (1993).

\bibitem{3b} E.~Bagan, H~G.~Dosch, P.~Gosdzinsky, S.~Narison and J.~M. Richard,
Z. Phys. C.~{\bf 64}, 57 (1994).

\bibitem{3} B.~L.~Ioffe, ArXiv hep-ph: 0810.4234.

\bibitem{4} C.~Adami and I.~Zahed, Phys. Rev. D~{\bf45}, 4312 (1992).

\bibitem{5} E.~G.~Drukarev, M.~G.~Ryskin, V.~A.~Sadovnikova, Prog. Part.
Nucl. Phys. {\bf47}, 73 (2001).

\bibitem{6} K.~G.~Wilson, Phys. Rev. {\bf179}, 1499 (1969).

\bibitem{6a} D.~Espiru, P.~Pascual and R.~Tarrach, Nucl. Phys. B {\bf 214},
285 (1983).

\bibitem{BL} B. L. Ioffe, Z. Phys. C~ {\bf 18}, 67 (1983).
\bibitem{R} L.~J.~Reinders, H.~Rubinstein, and S.~Yazaki, Phys. Rep.{\bf 127}, 1 (1987).

\bibitem{8} M. E. Peskin, Phys. Lett. B~{\bf88}, 126 (1979).

\bibitem{9} M. Gell-Mann, R. J. Oakes, and B.~Renner, Phys. Rev.
{\bf175}, 2195 (1968).

\bibitem{10} A. I. Vainshtein, V. I. Zakharov and M. A.~Shifman, Sov.
Phys. JETP Letters, {\bf27}, 50 (1978).

\bibitem{11} B. L. Ioffe and K. N. Zyablyuk, Eur. Phys. J. C~{\bf27}, 229
(2003);\\
K. N. Zyablyuk, JHEP {\bf0301}, 081 (2003);\\
A. V. Samsonov, hep-ph/047199.

\bibitem{12}
R. A. Bertlmann, C. A. Dominguez, M. Loewe, M. Perrotet and E. de
Rafael, Z. Phys. C.~{\bf 39}, 231 (1988),\\
B. V. Geshkenbein, Phys. Atom. Nucl. ~{\bf59}, 289 (1996).


\bibitem{BL1} B.~L.~Ioffe, Prog. Part. Nucl. Phys. {\bf 56}, 232
(2006).

\bibitem{13} V.~A.~Sadovnikova, E.~G.~Drukarev, and M.~G.~Ryskin, Phys.
Rev.  D~{\bf72}, 114015 (2005).

\bibitem{14} V.~A.~Sadovnikova, E.~G.~Drukarev, and M.~G.~Ryskin, Phys.
At. Nucl. {\bf71}, 1431 (2008); Yad. Fiz. {\bf71}, 1459 (2008).


\bibitem{BR1} R.~D.~Matheus, F.~S.~Navarra and M.~Nielsen, Braz. J. Phys. {\bf 36}, 1397 (2006).

\bibitem{BR2}R.~D.~Matheus, F.~S.~Navarra, M.~Nielsen and R.~Rodrigues da Silva, Phys. Rev. D
{\bf 76}, 056005 (2007).




\bibitem{RMF} Particle Data Group, Phys. Lett. B, {\bf 667}, 1 (2008).
\bibitem{15} M. Jamin, Z. Phys. C~{\bf37}, 635 (1988).

\bibitem{16} A.~A.~Ovchinnikov, A.~A.~Pivovarov, and L.~R.~Surguladze,
Int. J. Mod. Phys. A~{\bf6}, 2025 (1991).

\bibitem{17} D.~B.~Leinweber, Ann. Phys. (NY) {\bf254}, 328 (1997).

\bibitem{18} H.~Shiomi and T.~Hatsuda, Nucl. Phys. A~{\bf594}, 294
(1995).

\end{thebibliography}
\end{document}